\documentclass[preprint,12pt]{elsarticle}

\usepackage{amssymb}
\usepackage{amsmath}
\journal{Physics Letters A}

\begin{document}



\begin{frontmatter}



\title{Pilot-waves and copilot-particles: A nonstochastic approach to objective wavefunction collapse}

\author[label1]{Axel van de Walle}
\affiliation[label1]{organization={School of Engineering, Brown University},
            addressline={Box D},
            city={Providence},
            postcode={02912},
            state={RI},
            country={USA}}

%% Abstract
\begin{abstract}
We propose an extension to Schr\"odinger's equation that incorporates macroscopic measurement-induced wavefunction collapse. We find that a hybrid between two leading approaches, the Bohm-de Broglie pilot-wave and objective collapse theories, accomplishes this goal in accordance with Born's rule. We posit that the wavefunction guides the Bohmian particle and, conversely, the wavefunction gradually localizes towards the particle's position. As long as the particle can visit any state, as in a typical microscopic system, the localization effect does not favor any particular quantum state and, on average, the usual Schr\"odinger-like time evolution results. However, when the wavefunction develops spatially well-separated lobes, as would happen during a macroscopic measurement, the Bohmian particle can remain trapped in one lobe, which causes the wavefunction to eventually localizes. This loss of ergodicity mechanism recasts one of the foundational postulate of quantum mechanics as an emergent feature and has important implications regarding the feasibility of large-scale quantum computing.
\end{abstract}

\begin{graphicalabstract}
\includegraphics{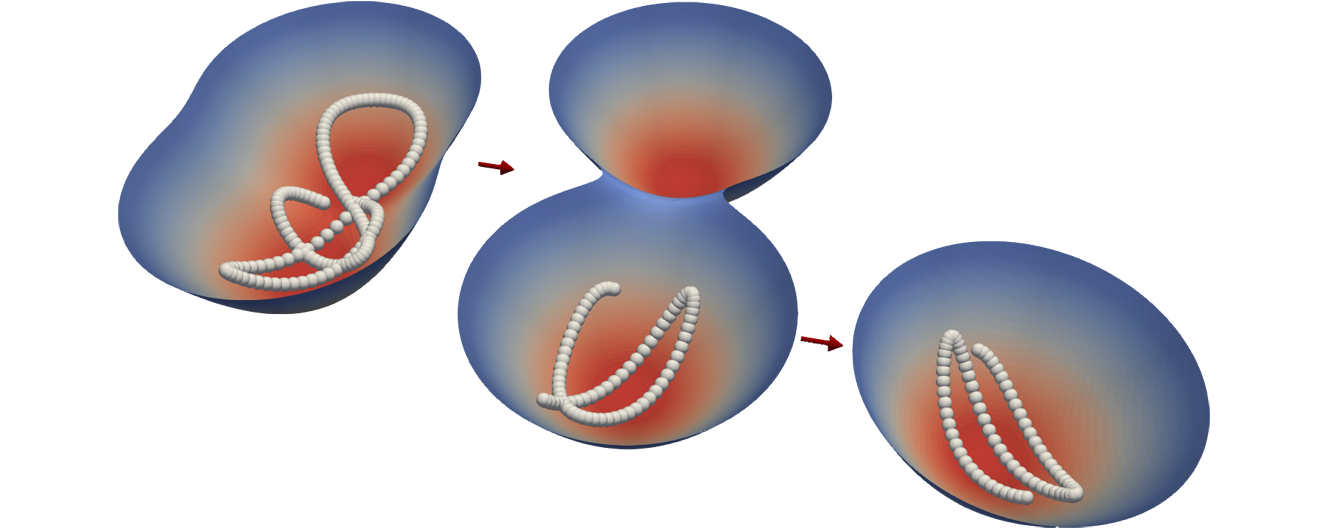}
\end{graphicalabstract}

\begin{highlights}
\item Hybrid pilot-wave and objective wavefunction collapse theory proposed.
\item Incorporates mutual feedback between Bohmian particle and wavefunction.
\item Dynamics reduces to Schr\"odinger equation for microscopic systems.
\item In macroscopic systems, loss of ergodicity mechanism yields wavefunction collapse.
\item Born rule naturally arises as a consequence rather than as a postulate.
\end{highlights}

%% Keywords
\begin{keyword}
Measurement postulate \sep Bohmian mechanics \sep Nonlinear wave equation
\end{keyword}

\end{frontmatter}

\section{Introduction}

Over a century after the birth of quantum mechanics,\ the search for a
compelling understanding of wavefunction \textquotedblleft
collapse\textquotedblright\ in quantum mechanics is still ongoing
\cite{cavalcanti:persp,wiseman:wigfri,bassi:exptest,spekkens:contex}. Existing
proposals include decoherence \cite{zurek:deco,schloss:deco}, pilot-waves or
Bohmian mechanics
\cite{valentini:bohmrev,bohm:hidvar12,broglie:pilot,durr:bohm}, many-world
interpretations \cite{carroll:many,everett:relative}, Bayesian interpretations
\cite{fuchs:qbism}, objective collapse \cite{bassi:revcollapse}, gravitational
effects \cite{penrose:gravqm,diodi:grav}, and other emerging ideas
\cite{schonfeld:cloud,balian:dynmeas,melkikh:nlqmeas,laloe:attract,barandes:sqcorr,wiseman:disguise}%
. These efforts have resulted in significant conceptual progress but no single
approach has been universally accepted as a solution to the measurement
problem in quantum mechanics \cite{schlosshauer:snapshot}.

As a step towards this goal, we propose a hybrid between Bohm-de Broglie
pilot-wave \cite{durr:bohm} and objective collapse theories
\cite{bassi:revcollapse}\ that incorporates a form of wavefunction collapse
consistent with Born's rule. Our approach aims to address some of the
conceptual questions associated with either approach: How do observers
perceive the Bohmian particle when a physical system's behavior only depends
on its wavefunction? What determines when and where the wavefunction
objectively collapses?

In analogy with pilot-wave theories, we posit that the global wavefunction
drives a Bohmian \textquotedblleft particle\textquotedblright. Our approach
however postulates a feedback in the other direction as well: The wavefunction
gradually localizes towards the particle's position. In the limit in which the
particle speed is high relative to the rate of localization, these
modifications to the Schr\"{o}dinger equation have little effect on the fairly
localized wavefunction of microscopic systems. But when the wavefunction
develops multiple macroscopically distinct regions of high probability
(hereafter, \textquotedblleft lobes\textquotedblright), the particle may
become trapped in one of these regions and the wavefunction starts to localize
towards only one of them. Since the probability of \textquotedblleft
selecting\textquotedblright\ one lobe is equal to the probability of the
particle lying in this lobe at the moment the different parts of the
wavefunction lose connectivity, Born's rule arises from this process as well.
Obtaining this trapping behavior in the macroscopic limit demands the use of
different equations of motion than those of conventional pilot-wave theory.

\section{Theory}

The theory is more naturally presented in the context of the Schr\"{o}dinger
equation for continuous degrees of freedom (see Appendix \ref{appdisc}\ for a
treatment of discrete states). Below, $\psi\left(  x,t\right)  $ is a properly
(anti-)symmetrized wavefunction in position representation, with $t$ being
time and $x$ being a $3N$-dimensional vector for a system of $N$ particles.
(We consider the familiar context of \textquotedblleft first
quantization\textquotedblright, but there is no fundamental problem associated
with applying it to boson fields within quantum field theory
\cite{struyve:bohmqft}, in which case $x$ would now be function-valued and
represent a specific value of a classical field over space.) Like most
theories with either a pilot-wave or an objective collapse flavor, our theory
is non-relativistic. Our proposed modified Schr\"{o}dinger equation takes the
form:%
\begin{equation}
\frac{d\psi\left(  x,t\right)  }{dt}=\frac{1}{\mathbf{i}\hbar}H\psi\left(
x,t\right)  -\kappa\psi\left(  x,t\right)  +\kappa\frac{\bar{\phi}_{r\left(
t\right)  }\left(  x\right)  \psi\left(  x,t\right)  }{\left\vert \psi\left(
r\left(  t\right)  ,t\right)  \right\vert ^{2}}\label{eqmodse}%
\end{equation}
where the Hamiltonian has the usual form $H=-\frac{\hbar^{2}}{2}m^{-1}%
\nabla^{2}+V\left(  x\right)  $ for a given imposed potential $V\left(
x\right)  $ and a given diagonal matrix of masses $m$. The first term of the
right-hand side of (\ref{eqmodse}) is standard while the remaining two
implement the localization of the wavefunction towards the particle located at
$r\left(  t\right)  $ (in $\mathbb{R}^{3N}$) at time $t$. The second term
introduces a tendency for exponential decay of the magnitude of the
wavefunction everywhere, without affecting its phase. The parameter $\kappa$
controls the rate of this process.\ In the last term, $\bar{\phi}_{r}\left(
x\right)  $ is a smooth function that only takes on a large value near the
position $r$ of the Bohmian particle: $\bar{\phi}_{r}\left(  x\right)
=\frac{1}{\#\Pi}\sum_{\pi\in\Pi}\sigma^{-3N}\phi\left(  (x-\pi r)/\sigma
\right)  $, where $\phi\left(  u\right)  $ is a localized function normalized
to integrate to $1$ (e.g. a standard Gaussian), $\sigma$ controls the spatial
extent of the localization and $\Pi$ is the set of coordinate permutations
among identical particles. This symmetrization device ensures that the
localization effect does not destroy the symmetry or anti-symmetry of the
wavefunction. Note that, since neither of the last two terms affect the phase
of the wavefunction, both Fermions and Bosons can be handled in the same way:
the sign alternation needed for Fermions is already embedded in the
wavefunction and there is no need for sign alternation in the $\bar{\phi}%
_{r}\left(  x\right)  $ term. We also implement the localization through a
smooth function $\phi\left(  u\right)  $ instead of, say, a delta function, to
avoid generating high-frequency noise in the wavefunction that, if present,
would likely have been experimentally detected from their associated
electromagnetic emissions \cite{bassi:exptest}. This consideration suggests
that $\sigma$ should not be much smaller than typical orbital sizes, so
$\sigma\gtrsim10^{-11}%
\operatorname{m}%
$.\ The net effect of the last term is an increase in wavefunction magnitude
near the particle. The normalization by $\left\vert \psi\left(  r,t\right)
\right\vert ^{2}$ ensures that, on average, the last two terms cancel out, as
we will demonstrate shortly.

There remains to specify the equation of motion for $r$. We depart from the
standard Bohm-de Broglie prescription and instead assume that the particle
experiences a fictitious potential $-\tau\ln\left\vert \psi\left(  r,t\right)
\right\vert ^{2}$, where $\tau$ is a constant, and otherwise obeys a classical
equation of motion:%
\begin{equation}
\frac{d^{2}r}{dt^{2}}=\mu^{-1}\tau\nabla_{r}\ln\left\vert \psi\left(
r,t-\Delta t\right)  \right\vert ^{2},\label{eqd2r}%
\end{equation}
where $\mu$ is a $3N\times3N$ diagonal matrix of fictitious masses. We allow
for a small time delay $\Delta t$ in the effect of the wavefunction (having
$\left\vert \psi\left(  r,t-\Delta t\right)  \right\vert $ instead of the
instantaneous effect $\left\vert \psi\left(  r,t\right)  \right\vert $) to
avoid a self-interaction artifact (to be discussed in more details below).
Standard statistical mechanics arguments \cite{mcquarrie:statmech}\ imply
that, in equilibrium at a fictitious temperature $\tau$ (expressed in energy
units), the particle's position $r$ will be distributed as $\exp\left(
-\left(  -\tau\ln\left\vert \psi\left(  r,t\right)  \right\vert ^{2}\right)
/\tau\right)  =\left\vert \psi\left(  r,t\right)  \right\vert ^{2}$, as
desired. Here, we assume that $\Delta t$ is much smaller than the time scale
of density evolution, so that $\left\vert \psi\left(  r,t-\Delta t\right)
\right\vert ^{2}\approx\left\vert \psi\left(  r,t\right)  \right\vert ^{2}$.
It is interesting that the logarithm in Equation (\ref{eqd2r}) coincidentally
accomplishes two tasks: separating out the magnitude from the phase (since
$\ln\left\vert \psi\left(  r,t\right)  \right\vert ^{2}=2\operatorname{Re}%
\ln\psi\left(  r,t\right)  )$ and placing the magnitude on a logarithmic
scale, as needed to cancel out the Boltzmann factor.

It is important to note that the $3N$ velocity degrees of freedom can play the
role of thermal bath, so no additional external thermal bath is needed to
invoke a thermalization argument. This presupposes that the velocities are
initialized (in the distant past) consistently with the fictitious temperature
$\tau$, e.g. $\frac{dr_{k}}{dt}\sim\operatorname{Normal}\left(  0,\mu_{k}%
^{-1}\tau\right)  $ for $k=1,\ldots,3N$. This argument also assumes that the
particle equilibrates rapidly relative to the time scale of changes in the
wavefunction magnitude $\left\vert \psi\left(  r,t\right)  \right\vert $. This
equilibration behavior is implied by Boltzmann's classic \textquotedblleft
H-theorem\textquotedblright, although it is difficult to formulate simple
general conditions guaranteeing a given rate of convergence (for a recent
account of these types of results, see \cite{villani:hthm}). In fact, in the
sequel we assume that such fast equilibration behavior is guaranteed only for
microscopic systems. A simple dimensional argument indicates that the
equilibration (or relaxation) time $t_{r}$ should scale with $\ell/\sqrt
{\mu^{-1}\tau}$, where $\ell$ is of the order of magnitude of the linear
spatial extent of a particle's wavefunction while $\sqrt{\mu^{-1}\tau}$ is the
average velocity of the fictitious particle in thermal equilibrium. Note that,
from now on, we take $\mu$ to be a multiple of the identity matrix and write
it as a scalar, for simplicity.

Note that all quantities labelled here as \textquotedblleft
fictitious\textquotedblright\ are not tied to any real observable properties
of the system: $\mu$ is not necessarily related to the actual particles'
masses and $\tau$ may not be the real temperature. In fact, these quantities
enter all equations as the ratio $\mu^{-1}\tau$, which we leave as an
adjustable parameter in the theory that controls the equilibration time. A
simple order-of-magnitude calculation is helpful to set the scale of these
constants. A typical atomic orbital is of the order of $\ell\approx10^{-9}%
\operatorname{m}%
$, while atomic vibrations have a typical period of the order of $10^{-14}%
\operatorname{s}%
$, which estimates the maximum allowed relaxation time $t_{r}$ that would not
significantly affect atomic dynamics. If we assume that the fictitious
particle would take about $n_{eq}\approx1000$ trips across the orbital to
equilibrate, we then have that the fictitious particle velocity would need to
be about $\sqrt{\mu^{-1}\tau}\approx n_{eq}\ell/t_{r}=10^{8}%
\operatorname{m}%
/%
\operatorname{s}%
$, which is comfortably below the speed of light.%

\begin{figure}
\centerline{\includegraphics[width=2.8in]{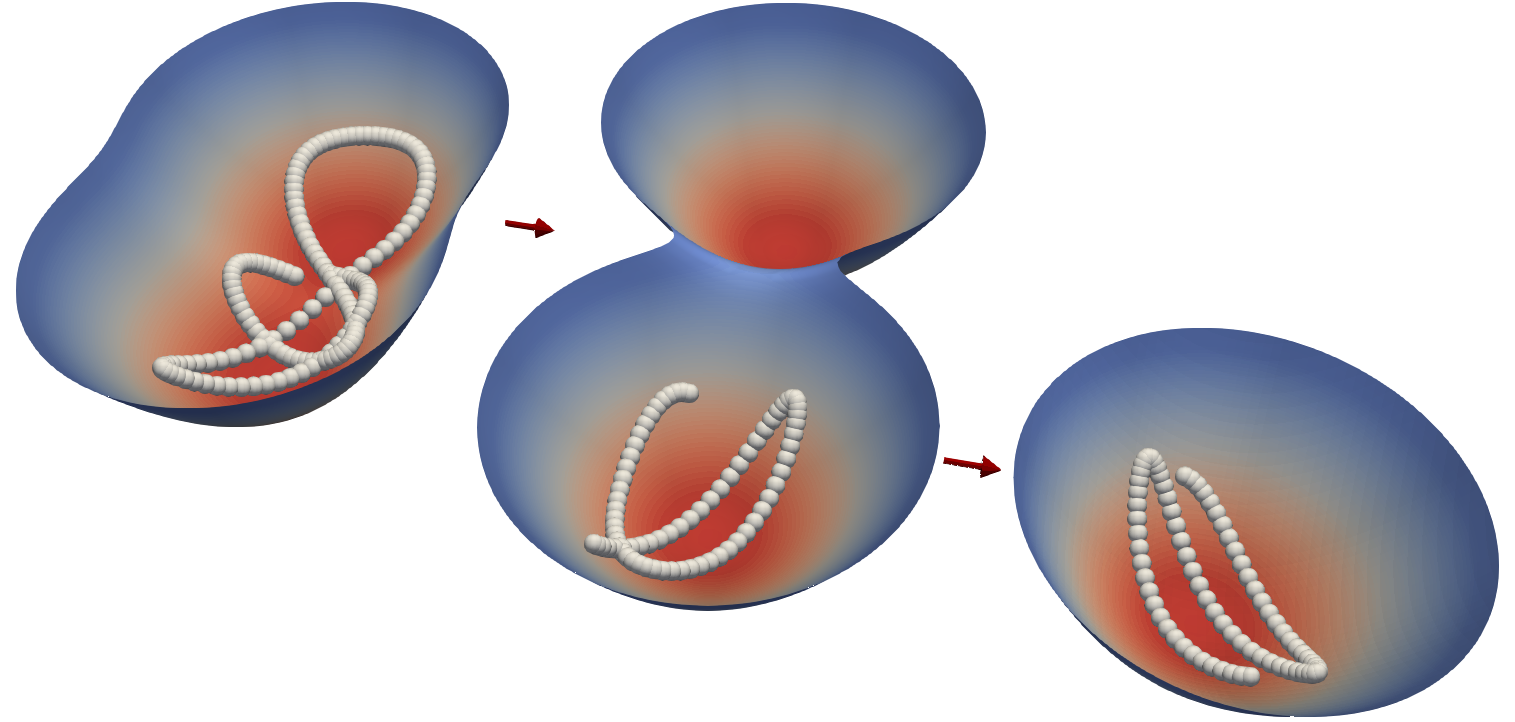}}
\caption{(Color online) Microscopic vs.\ macroscopic behaviors. Upper left:
For a spatially localized wavefunction, the Bohmian particle
(represented by a  grey spheres trail) can access all regions where the
wavefunction is nonnegligible (here represented by a fictitious potential
surface proportional to $- \ln| \psi(r,t)|^2$).
Middle: As a superposition of states starts to involve spatially well-separated states,
the particle remains trapped in one of the ``lobes'' of the wavefunction.
Lower right: Loss of ergodicity causes wavefunction localization around one of the
macroscopically distinct outcomes.}
\label{fig:split}
\end{figure}%

Traditional Bohm-de Broglie dynamics guarantees a strong equivariance
property: If an ensemble of Bohmian particles is distributed according to
$\left\vert \psi\left(  r,t_{0}\right)  \right\vert ^{2}$ at some time $t_{0}%
$, it will remain, by construction, distributed according to $\left\vert
\psi\left(  r,t\right)  \right\vert ^{2}$ for all $t>t_{0}$. However, we
deviate from standard Bohm-de Broglie dynamics for two reasons. First, the
particles tend to not get trapped very well under the usual Bohm-de Broglie
dynamics: As the particle approaches a region of low probability density, it
typically speeds up (which specifically ensures that the particle has low
probability of being found there). As a result, the particle can easily escape
any isolated lobe in the wavefunction. Second, the Bohmian particle does not
thermalize very quickly. For instance, in a wave packet, the particle
typically follows along smoothly while remaining at the same location relative
to the center of the wave packet, instead of visiting all positions of high
probability. Our alternative dynamics avoids both of these problems. The
equivariance property under Bohm-de Broglie dynamics is here replaced by a
convergence towards $\left\vert \psi\left(  r,t\right)  \right\vert ^{2}$
implied by Boltzmann's \textquotedblleft H-theorem\textquotedblright%
\ \cite{villani:hthm}. When the relaxation time associated with this process
becomes very long, the particle's distribution could differ from $\left\vert
\psi\left(  r,t\right)  \right\vert ^{2}$ and this possibility will in fact
turn out to be essential to enable wavefunction collapse in macroscopic
systems in our formalism.

We are now ready to show that, under conditions ensuring rapid thermalization
(i.e. when the particle's relaxation time $t_{r}$ is sufficiently short), our
proposal essentially reduces to the conventional Schr\"{o}dinger equation.
Indeed, computing the expected value of the last term of (\ref{eqmodse}) with
respect to the random position $r$ with instantaneous probability density
$\left\vert \psi\left(  r,t\right)  \right\vert ^{2}$, we have:%
\begin{align}
&  \int_{r\in\mathbb{R}^{3N}}\kappa\bar{\phi}_{r}\left(  x\right)  \frac
{1}{\left\vert \psi\left(  r,t\right)  \right\vert ^{2}}\left\vert \psi\left(
r,t\right)  \right\vert ^{2}\psi\left(  x,t\right)  ~dr\nonumber\\
&  =\kappa\frac{1}{\#\Pi}\sum_{\pi\in\Pi}\int_{r\in\mathbb{R}^{3N}}%
\sigma^{-3N}\phi\left(  \left(  x-\pi r\right)  /\sigma\right)  dr~\psi\left(
x,t\right)  =\kappa\psi\left(  x,t\right)  \label{eqvschro}%
\end{align}
since $\int\phi\left(  x-\pi r\right)  dr=1$ by construction. As a result, on
average, the last term cancels out the second term of (\ref{eqmodse}) and we
recover Schr\"{o}dinger's equation.

In contrast, as illustrated in Figure \ref{fig:split}, if the wavefunction
develops spatially well-separated lobes and the particle is trapped in only
one of them (i.e. when the particle's global relaxation time becomes very
long), then the last term remains essentially zero for the lobes that are no
longer visited and, there, only the decaying term $-\kappa\psi\left(
x,t\right)  $ is active, resulting in an exponential decay of the
wavefunction. Within the lobe trapping the particle, the localization term of
(\ref{eqmodse}) is larger than average, since the particle visits points
therein more often than would be indicated by the wavefunction magnitude. The
end result is that the localization term there exceeds the decaying term on
average and the wavefunction gradually concentrates onto that lobe at the
expense of the others. The loss of ergodicity of the Bohmian particle is the
source of the wavefunction collapse within this model. The idea that
non-equilibrated Bohmian particles can lead to deviations from standard
quantum mechanics has precedents \cite{valentini:noneq}, but our proposed
dynamics differ form these earlier proposals and lead to a clear mechanism for
objective collapse. This helps explain single outcomes in macroscopic
measurements, since they\ arguably lead to wavefunctions with macroscopically
disconnected lobes, such as a dial (or \textquotedblleft
pointer\textquotedblright) being in one place or another or a photon being
emitted in different directions, etc. Appendix \ref{appsubsys} extends this
reasoning to the measurement of a subsystem.

Another mechanism that could cause a loss of ergodicity is the process of
copying a microscopic state onto a large number of particles (within the
limits of the no-cloning theorem). This operation typically results in
increasingly deeper basins in the fictitious potential $-\tau\ln\left\vert
\psi\left(  r,t\right)  \right\vert ^{2}$ that increase the likelihood of the
particle being trapped. To see this, consider a particle in a superposition of
two states $\alpha\left\vert 1\right\rangle +\beta\left\vert 2\right\rangle $
(with $\left\vert \alpha\right\vert ^{2}+\left\vert \beta\right\vert ^{2}=1$)
that is starting to interact with a measurement apparatus made of $N-1$
distinguishable particles, initially in a \textquotedblleft no
measurement\textquotedblright\ state $\left\vert 0\right\rangle \otimes
\cdots\otimes\left\vert 0\right\rangle $. After a suitable unitary evolution,
the joint system transitions to the state $\alpha\left\vert 1\right\rangle
\otimes\left\vert 1\right\rangle \otimes\cdots\otimes\left\vert 1\right\rangle
+\beta\left\vert 2\right\rangle \otimes\left\vert 2\right\rangle \otimes
\cdots\otimes\left\vert 2\right\rangle $. Now assume that, for particle $i$,
the states $\left\vert 1\right\rangle $ and $\left\vert 2\right\rangle $ are,
respectively, associated with single-particle wavefunctions in position
representation $\psi_{1}\left(  x_{i}\right)  =\varphi\left(  x_{i}-a\right)
$ and $\psi_{2}\left(  x_{i}\right)  =\varphi\left(  x_{i}+a\right)  $ for
some macroscopic value $a\in\mathbb{R}^{3}$ and centrosymmetric unimodal
function $\varphi\left(  x_{i}\right)  $ (with $x_{i}\in\mathbb{R}^{3}$). The
overall wavefunction is then%
\[
\psi\left(  x\right)  =\alpha\prod\nolimits_{i=1}^{N}\varphi\left(
x_{i}-a\right)  +\beta\prod\nolimits_{i=1}^{N}\varphi\left(  x_{i}+a\right)  .
\]
To estimate the escape rate out of one well, we calculate the fictitious
energy barrier $\Delta\epsilon$, defined as the difference between the
particle's fictitious energy at the saddle point at $x_{i}=0$ (for all $i$)
and at the bottom of one of the wells (say, at $x_{i}=a$, for all $i$):
$\Delta\epsilon\equiv C+N2\tau\ln\left(  \left\vert \varphi\left(  a\right)
\right\vert /\left\vert \varphi\left(  0\right)  \right\vert \right)  $, where
we assumed $\varphi\left(  2a\right)  \approx0$ and $C$ is a constant
independent of $N$. Since the barrier $\Delta\epsilon$ increases linearly with
$N$, conventional transition state theory \cite{vineyard:tst}\ implies that
the transition rate between the two wells is proportional to $\exp\left(
-\Delta\epsilon/\tau\right)  \propto\left\vert \varphi\left(  0\right)
/\varphi\left(  a\right)  \right\vert ^{2N}$\ and thus decreases exponentially
with the number $N$ of particles involved in the measurement.

\begin{figure}
\centerline{\includegraphics[width=3.2in]{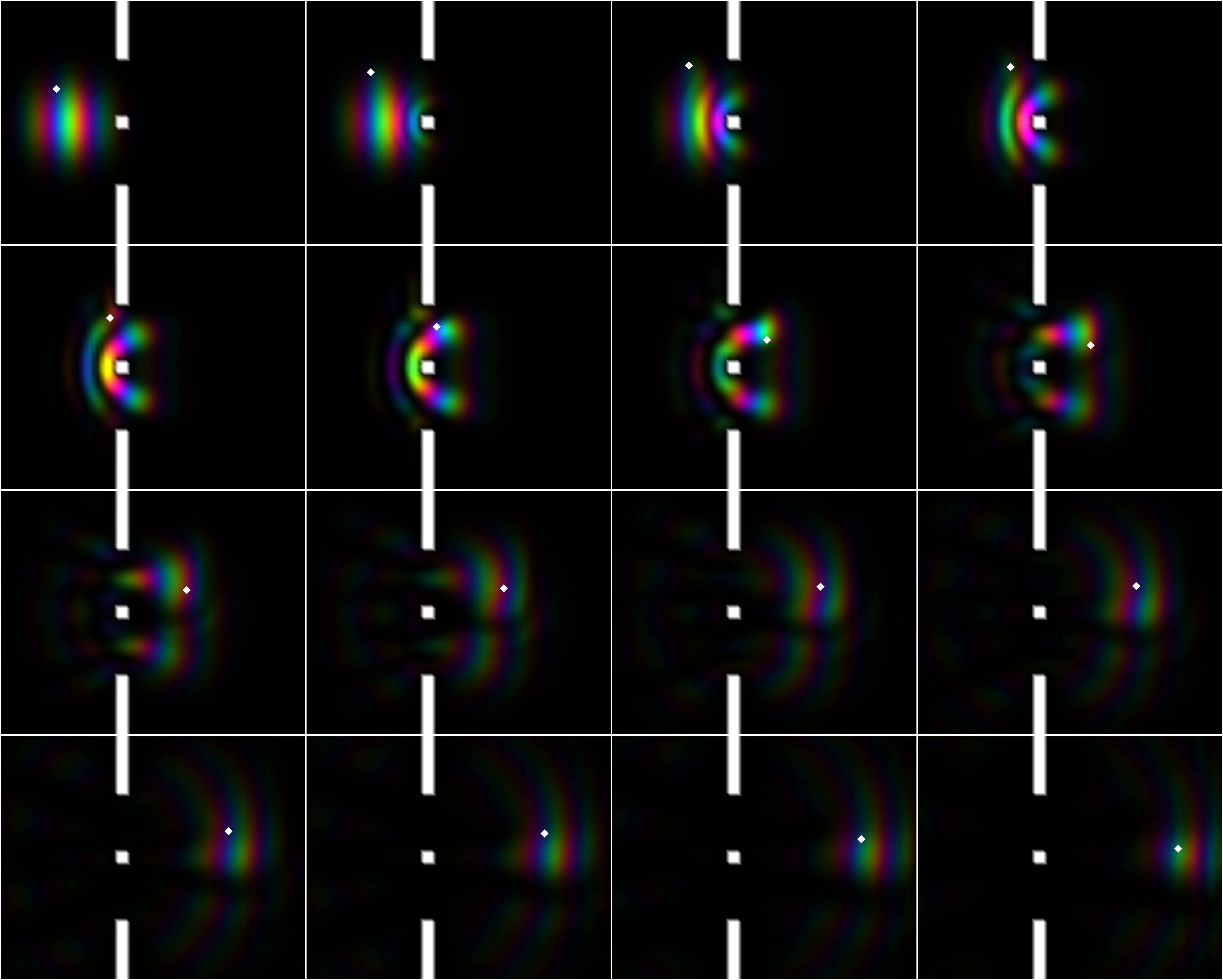}}
\caption{(Color online) Snapshots of double-slit experiment simulation with
modified Schr\"{o}dinger dynamics.
Wavefunction magnitude indicated by intensity and phase by color.
Position of the Bohmian particle marked by white lozenge.
(Simulation parameters: electron particle, 7.5 nm $\times
$ 6.0 nm area, 0.08 fs duration,
$\kappa=2\times10^{15}$ s$^{-1}$, $\sigma=0.02$ nm,
$\tau/\mu=5\times10^{13}$ m$^{2}$/s$^{2}.$)
Magnitude of the decay rate parameter $\kappa$
exaggerated to show wavefunction collapse over a short simulation time.}
\label{fig:ds_tiled}
\end{figure}

Figure \ref{fig:ds_tiled} illustrates our proposed time evolution of the
wavefunction in a classic setting: the double-slit experiment.\ The simulation
shows both the initial wave-like interference behavior and the eventual
collapse to a more specific outcome as the wave packet heads towards a
detector. The Supplementary materials include an animation of the same
numerical experiment, as well as the simulation code itself. Appendix
\ref{appepr}\ also includes a treatment of a standard EPR-type experiment
\cite{EPR:org}, while Appendix \ref{apptrans} explores the transition from
short to long equilibration times via numerical and analytic means.%

\section{Discussion}

The proposed theory improves upon the Bohm-de Broglie pilot-wave approach
along a few dimensions. First, it avoids the awkwardly distinct treatment of
(i) the physical system being described by the wavefunction and (ii) the
observer only perceiving the position of the Bohmian particle. Having a
feedback mechanism through which the particle affects the wavefunction solves
this problem and also leads to a model of wavefunction collapse. In our
approach, both the wavefunction and the particle jointly describe the behavior
of both the physical system and the observer. The definite state of the
particle is what drives the wavefunction localization during the measurement
process. A conceptually appealing feature of the theory is that it avoids the
often criticized exponential growth of alternative realities in which the
Bohmian particle does not reside and that seem entirely superfluous to
describe the actual state of the world. In our approach, the wavefunctions
associated with these alternative realities simply die out exponentially fast.

The proposed theory also improves some aspects of objective collapse theories,
in which the position of the collapse is simply picked at random based on the
wavefunction and a random noise field. On a conceptual level, our approach
provides a rationale for the origin of this randomness: The random location of
the collapse is simply determined by the random location of the particle. The
type of collapse we propose may also be more amenable to an eventual
relativistic treatment, as we shall discuss further below. On a more empirical
level, a problem with objective collapse theories is the abruptness of the
collapse process, which is associated with a high-frequency signature which
has yet to be found experimentally \cite{bassi:exptest}. In our approach, the
collapse is a smoother process more consistent with experimental tests, which
distinguishes it from some other proposals aimed at combining Bohmian and
collapse theories \cite{zanghi:bm_grw}. Collapse theories that employ a
\textquotedblleft colored\textquotedblright\ noise can also lead to a smoother
collapse and some can be cast in a Bohmian framework \cite{wiseman:disguise},
although, in this approach, the Bohmian variables are introduced only for the
environment (consisting of harmonic oscillators), while our approach treats
the environment and the system of interest on an equal footing.

The idea that the wavefunction is attracted to particles\ can be found in
earlier work \cite{laloe:attract,darrow:2way}, but our approach differs in
substantial ways. Unlike \cite{darrow:2way}, we show that the two-way coupling
leads to Born's rule. Unlike \cite{laloe:attract}, we consider collapse
towards a point in $\mathbb{R}^{3N}$ instead of towards a single-particle
density, which helps maintain EPR-type correlations between distant
measurements. Also, to ensure particle trapping, our particle dynamics differs
from those of both \cite{darrow:2way} and \cite{laloe:attract}.

The proposed theory is compatible with the standard decoherence picture (see
Appendix\ \ref{appdens}\ for a density matrix formulation): it does not affect
the basic idea that interactions with the environment cause a decay of the
off-diagonal terms of a subsystem's density matrix. But once the resulting
mixture of outcomes involves macroscopically distinct states, our equation of
motion predicts localization towards one of these macroscopic states. Hence,
it provides an objective \textquotedblleft selection rule\textquotedblright%
\ that is absent under decoherence alone and avoids the need for a many-world
interpretation. The decoherence concept makes it unnecessary for the
wavefunction to collapse rapidly in our framework. The conversion from a
superposition of states to a statistical mixture takes place over the same
time span under our theory as under decoherence alone. The exponential decay
of all but one branch of the statistical mixture, however, can take much
longer: any decay rate faster than human perception (i.e., milliseconds) would
be sufficient to give human observers the experience of a single measurement
outcome, which suggests that $\kappa$ is no smaller than $10^{3}%
\operatorname{s}%
^{-1}$. At the same time, $\kappa$ should not be such that the wavefunction
could decay before the fictitious particle has had time to equilibrate. Taking
our prior estimate of $t_{r}\approx10^{-14}%
\operatorname{s}%
$ places a rough upper bound of $10^{14}%
\operatorname{s}%
^{-1}$ on $\kappa$.

The proposed theory does exhibit some undesirable features (see also Appendix
\ref{appobj}). Most obviously, the proposed modified Schr\"{o}dinger equation
is slightly nonlinear, whereas linearity of quantum mechanics has been
verified experimentally to a great precision \cite{bollinger:linqm}. This
suggests that parameter $\kappa$ should be closer to our lower bound than to
the upper bound. The fact that the nonlinearity must be small mitigates other
known issues associated with nonlinearity \cite{polchinski:nlqmbug}. The most
troubling of these issues is the possibility of faster-than-light
communication. However, the objective collapse literature
\cite{bassi:objcoluni} already offers a natural solution to this possible
problem that is relevant within our approach as well. Theorem 1 in
\cite{bassi:objcoluni} provides conditions under which faster-than-light
communication is precluded by a suitable combination of nonlinearity
\emph{with randomness}. The two conditions are that (i) the time evolution of
the density matrix is Markovian and (ii) nonlinearity and randomness cancel
out on average, so that the time evolution of the expected density matrix
remains a linear operator. This condition is exactly satisfied in objective
collapse theories. In Appendix \ref{appftl}, we derive conditions under which
Theorem 1 in \cite{bassi:objcoluni} applies to our model as well.

While our current treatment is non-relativistic, many aspects of our framework
are amenable to a relativistic extension. The collapse phenomenon involves
local operations: the decay term $-\kappa\psi\left(  x,t\right)  $ is purely
local while the localization term only involves instantaneous effects over
short distances (as governed by the function $\phi_{r}\left(  x\right)  $) and
ensuring that this effect propagates at a finite speed entails no significant
changes in the theory. Since relativistic versions of pilot-wave
\cite{zanghi:bohmrel,darrow:2way} and objective collapse theories
\cite{tumulka:grwrel}\ have been developed with some success, similar ideas
could carry over in the present setup. While the presence of a global hidden
variable (the particle's position $r$) could in principle clash with
relativity, there are compelling reasons to conclude that it does not, as
detailed in Appendix \ref{appftl}.

In our theory, the time delay $\Delta t$ should satisfy\ $\Delta t\gg
\sigma/\sqrt{\tau/\mu}$ (with scalar mass $\mu$ for simplicity) to avoid a
strong instantaneous self-interaction. In this fashion, the wavefunction
localization only starts to have an effect on the particle when it has already
moved by $\Delta t$ times its average velocity $\sqrt{\tau/\mu}$. If this
distance exceeds $\sigma$, the particle will not experience the
\textquotedblleft indentation\textquotedblright\ it has just created in the
wavefunction. Absent this delay, the particle would have a tendency to
gradually lose kinetic energy. On a physical level, the time delay
incorporates the fact that the wavefunction might take some time to be fully
\textquotedblleft indented\textquotedblright\ by the particle. A more exact
theory might have a general spatio-temporal interaction kernel that gradually
goes from no interaction to full interaction over a period of the order of
$\Delta t$. One could formulate general conditions on the time and spatial
derivatives of that kernel that ensure that the particle does not experience a
net drag. Here, we simply choose to approximate the time-dependence of this
hard-to-determine kernel by a delta function shifted by a lag of $\Delta t$,
since its exact shape has little impact on observables.

A time delay is not the only way to address the self-interaction problem: one
could add a negative drag term in the equation of motion for $r$ instead, or
place the particle in weak interaction with an outside reservoir that
constantly replenishes the lost energy to maintain an average kinetic energy
$\tau$. We favor the time delay approach for its simplicity, but it would be
difficult to discern these alternatives experimentally. To pick up the
relevant time history of $r\left(  t\right)  $, the time delay $\Delta t$
should also be far less than the relaxation time of the fictitious particle,
which we had bounded by $t_{r}\lesssim10^{-14}%
\operatorname{s}%
$. Combined with our estimate of $\sqrt{\tau/\mu}\approx10^{8}%
\operatorname{m}%
/%
\operatorname{s}%
$, this implies that $\sigma\ll t_{r}\sqrt{\tau/\mu}\approx10^{-6}%
\operatorname{m}%
$.

\section{Conclusion}

The key new aspects of the theory include: (i) a mutual feedback between the
Bohmian particle and the wavefunction; (ii) a different proposed equation of
motion for the Bohmian particle; (iii) a mechanism based on the loss of
ergodicity (or trapping) of the Bohmian particle to explain wavefunction
collapse phenomena; (iv) the idea that the collapse mechanism can actually be
gradual and local. An interesting feature of our approach is that it explores
slight relaxations of some properties that are strictly imposed in other
theories. The equivariance of the particle's distribution found in Bohmian-de
Broglie pilot wave theories is replaced by fast convergence towards an
equilibrium distribution in microscopic systems while macroscopic systems can
experience loss of\ ergodicity. The strict prohibition of faster-than-light
communication in standard quantum mechanics and objective collapse theories is
relaxed so that superluminal features are in principle possible, but
constrained to small and difficult-to-access regimes. Finally, the strict
linearity of quantum mechanics is slightly relaxed and this fact could enable
tests of the proposed theory.

There already exists some suggestive, but not yet conclusive, evidence in
support of our proposal. For instance, it has been noted
\cite{schonfeld:cloud} that cloud chamber tracks do not originate at the
radioactive source itself, but rather a few centimeters away from it. This
observation suggests that wavefunction collapse takes place over time and
depends on spatial separation, although it could also be explained by a
velocity-dependent cross section for droplet nucleation, for instance.

Perhaps the most direct test would involve generating particles in a
superposition of two spatially-separated wavepacket-like states that are later
recombined, and then analyzing how contrast of the resulting interference
patterns depend on the spatial separation of the initial wavepackets. Very
sensitive experiments of this kind have been performed
\cite{kovachy:qmetrescale}, whose primary finding was to demonstrate that
interference contrast can survive even for wavepacket separated by about half
a meter for about a second. However, as an ancillary benefit, such experiments
also revealed a strong dependence of that contrast on the wavepacket
separation (see Figure 4b in \cite{kovachy:qmetrescale}). This observation is
consistent with our proposed mechanism but not entirely conclusive, as it
could also arise for other reasons, such as decoherence becoming more
pronounced for more distant wavepackets. Hence, a rigorous test would involve
comparing processes that are expected to undergo a similar amount of
decoherence but that differ in the length and/or time scale involved. Such
confirmation would shed light on some of the conceptual foundations of quantum
mechanics and could inform us regarding the feasibility of large-scale quantum computers.%

\clearpage

\appendix

\section{EPR-type experiments}

\label{appepr}It is instructive to see how the theory handles EPR-type
experiments \cite{EPR:org}. In this case, the long-distance correlations in
the wavefunctions simply induce similar long-distance correlations between
different components of the $3N$-dimensional particle position. As each
observer of one particle of the EPR pair makes a measurement, they each create
disjoint lobes in their wavefunction. Specifically, consider an EPR pair in a
state $\left(  \left\vert H\right\rangle \otimes\left\vert H\right\rangle
+\left\vert V\right\rangle \otimes\left\vert V\right\rangle \right)
\allowbreak/\sqrt{2}$ and one observer measures the first particle in the
basis $\left\vert H\right\rangle ,\left\vert V\right\rangle $ (e.g.,
horizontally or vertically polarized photon), while the other observer
measures the second particle in the basis $\left\vert P\right\rangle
=\cos\left(  \theta\right)  \left\vert H\right\rangle +\sin\left(
\theta\right)  \left\vert V\right\rangle $, $\left\vert M\right\rangle
=-\sin\left(  \theta\right)  \left\vert H\right\rangle +\cos\left(
\theta\right)  \left\vert V\right\rangle $. Expressed in the observers' basis
the pair's state is
\begin{align*}
&  \frac{\cos\left(  \theta\right)  }{\sqrt{2}}\left\vert H\right\rangle
\otimes\left\vert P\right\rangle -\frac{\sin\left(  \theta\right)  }{\sqrt{2}%
}\left\vert H\right\rangle \otimes\left\vert M\right\rangle \\
&  +\frac{\sin\left(  \theta\right)  }{\sqrt{2}}\left\vert V\right\rangle
\otimes\left\vert P\right\rangle +\frac{\cos\left(  \theta\right)  }{\sqrt{2}%
}\left\vert V\right\rangle \otimes\left\vert M\right\rangle .
\end{align*}
Now, we consider a simple measurement process that destructively copies the
state of the pair onto two macroscopic pointers (initially in some neutral
state $\left\vert \psi_{10}\right\rangle \otimes\left\vert \psi_{20}%
\right\rangle $ ). Specifically, the measurement process maps, for particle
$p=1,2$, the state $\left\vert s\right\rangle \otimes\left\vert \psi
_{p0}\right\rangle $ (for $s=H,V,P,M$) onto the state $\left\vert
0\right\rangle \otimes\left\vert \psi_{ps}\right\rangle $, where $\left\vert
0\right\rangle $ is some arbitrarily chosen neutral state (it could be any one
of $H,V,P,M$ --- as long as it is always the same state regardless of the
measurement outcome). (As an alternative, we could also consider a measurement
that maps $\left\vert s\right\rangle \otimes\left\vert \psi_{0}\right\rangle $
onto the state $\left\vert s\right\rangle \otimes\left\vert \psi
_{s}\right\rangle $, at the expense of necessitating an additional tracing out operation.)

After the measurement, the system and the pointers are in the state%
\begin{align*}
&  \left\vert 0\right\rangle \otimes\left\vert 0\right\rangle \otimes\left(
\frac{\cos\left(  \theta\right)  }{\sqrt{2}}\left\vert \psi_{1H}\right\rangle
\otimes\left\vert \psi_{2P}\right\rangle -\frac{\sin\left(  \theta\right)
}{\sqrt{2}}\left\vert \psi_{1H}\right\rangle \otimes\left\vert \psi
_{2M}\right\rangle \right.  \\
&  \left.  +\frac{\sin\left(  \theta\right)  }{\sqrt{2}}\left\vert \psi
_{1V}\right\rangle \otimes\left\vert \psi_{2P}\right\rangle +\frac{\cos\left(
\theta\right)  }{\sqrt{2}}\left\vert \psi_{1V}\right\rangle \otimes\left\vert
\psi_{2M}\right\rangle \right)
\end{align*}
whose probability density, expressed in the position basis for the pointers,
is:%
\begin{align*}
&  \frac{\cos^{2}\left(  \theta\right)  }{2}\left\vert \psi_{1H}\left(
x_{1}\right)  \right\vert ^{2}\left\vert \psi_{2P}\left(  x_{2}\right)
\right\vert ^{2}\\
&  +\frac{\sin^{2}\left(  \theta\right)  }{2}\left\vert \psi_{1H}\left(
x_{1}\right)  \right\vert ^{2}\left\vert \psi_{2M}\left(  x_{2}\right)
\right\vert ^{2}\\
&  +\frac{\sin^{2}\left(  \theta\right)  }{2}\left\vert \psi_{1V}\left(
x_{1}\right)  \right\vert ^{2}\left\vert \psi_{2P}\left(  x_{2}\right)
\right\vert ^{2}\\
&  +\frac{\cos^{2}\left(  \theta\right)  }{2}\left\vert \psi_{1V}\left(
x_{1}\right)  \right\vert ^{2}\left\vert \psi_{2M}\left(  x_{2}\right)
\right\vert ^{2},
\end{align*}
where we have assumed negligible overlap between the wavefunctions associated
with different pointer positions ($\psi_{ps}\left(  x\right)  \psi
_{ps^{\prime}}\left(  x\right)  \approx0$ for $s\not =s^{\prime}$). As these 4
lobes of probability have little spatial overlap, the Bohmian particle would
get trapped in one of them, with a probability equal to that particle being
found in that lobe at the time they became disconnected. The end results are
measurements with outcomes $\left(  H,P\right)  ,$ $\left(  H,M\right)  ,$
$\left(  V,P\right)  ,$ $\left(  V,M\right)  $ with probability $\cos
^{2}\left(  \theta\right)  /2,\sin^{2}\left(  \theta\right)  /2,\sin
^{2}\left(  \theta\right)  /2,\cos^{2}\left(  \theta\right)  /2$,
respectively, in agreement with Born's rule and the known outcome of EPR-type
experiments. Note that there are no contradictions with Bell inequalities
because the hidden variables $r$ are not local.

\section{Subsystem Measurements}

\label{appsubsys}
\begin{figure}
\centerline{\includegraphics{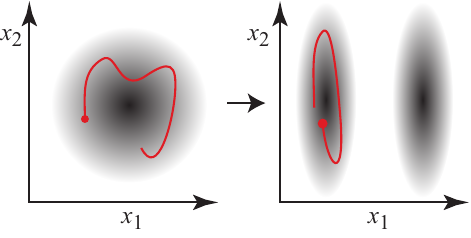}}
\caption
{(Color online) Schematic representation of the measurement of the position $x_1$ of one
particle, while other degrees of freedom (represented by $x_2$) are left unaffected.
Connectivity is lost only along the $x_1$ dimension.
(Particles are considered distinguishable for simplicity.)}
\label{fig:measpart}
\end{figure}

It is instructive to consider how the theory handles the fact
that one typically seeks to measure the state of only some subsystem and not
of the universe as a whole. In Figure \ref{fig:measpart}, the $x_{1}$
coordinate describes the degree of freedom to be measured, while the $x_{2}$
coordinate represents the rest of the universe. The degrees of freedom that
are being measured are simply those along which the wavefunction develops
disconnected lobes. The remaining degrees of freedom remain fully connected
and the particle can continue to reach its equilibrium distribution along
those dimensions. The net effect is that the remaining degrees of freedom
still closely follow Schr\"{o}dinger's equation and only the degrees of
freedom that are being measured experience significant deviations from unitary
evolution. A similar argument can be made for \textquotedblleft
weak\textquotedblright\ measurements.%

\section{Equilibration Behavior}

\label{apptrans}Here we present a numerical experiment that specifically
probes the transition from short to long equilibration times. We consider two
multidimensional localized wavefunction lobes ($\sim1$ nm in width)\ separated
by a given distance $d$ and observe the particle's equilibration behavior as a
function of separation $d$ (this setup is shown schematically in
Figure\ \ref{fig:split}). We monitor the positions visited by the particle
projected on the axis parallel to the line connecting the centers of the two
lobes and construct an associated histogram. The proximity to equilibrium is
quantified by an equilibrium index defined as the normalized inner product
between this histogram and the histogram expected in equilibrium (as
determined by fully converged Monte Carlo simulations).

Implementation details: The lobes are represented by the functional form
$\left\vert \psi\left(  r\right)  \right\vert ^{2}=e^{-\left\Vert \Lambda
^{-1}(r+ud/2)\right\Vert }+e^{-\left\Vert \Lambda^{-1}(r-ud/2)\right\Vert }$,
where $d$ is the lobe separation, $r$ is a vector in $\mathbb{R}^{32}$ and
$\Lambda$ is a conformable diagonal matrix with entries close to $1$ nm but
each perturbed differently by at most $0.05$ nm (this functional form is
inspired by the form of lowest energy $s$ atomic orbitals). The unit vector
$u$ essentially aligns with the first Cartesian axis, apart from deviations of
up to $0.1$ on average along the other axes. These various symmetry breakings
ensure that the test system is representative of a typical system rather than
an artificially symmetric one. We set $\sqrt{\tau/\mu}=7\times10^{6}$ m/s. The
reference Monte Carlo simulations employ the Metropolis algorithm with
Gaussian trial jumps of width $0.1$ nm as well as sign flips of the first
coordinate and perform 100,000 equilibration steps and 500,000 averaging steps.%

\begin{figure}
\centerline{\includegraphics{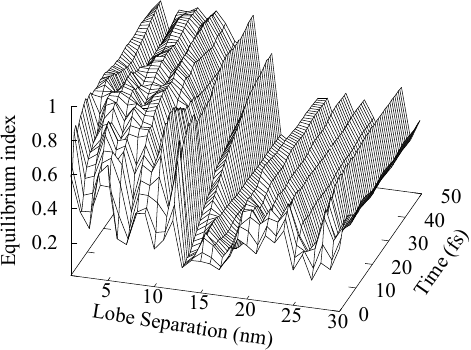}}
\caption{Equilibration behavior as a function of lobe separation and time.}
\label{fig:ergotest}
\end{figure}%

As seen in Figure \ref{fig:ergotest}, at short separations ($d\lesssim12$ nm),
the particle distribution converges to equilibrium (as indicated by an
equilibrium index approaching $1$) within about $10$ femtoseconds as the
particle regularly visits the two lobes. For $d\gtrsim12$ nm, there is a sharp
change in behavior: the equilibrium index remains low, indicating that the
system no longer reaches an equilibrium in which both lobes are visited. It is
only in a very narrow range of separations (around $d\approx12$ nm) that the
intermediate regime could in principle be visible. The exact location
($\sim12$ nm) of the transition depends on our specific simulation parameters,
but the sharpness of the transition does not, thus demonstrating that it is
only under rare circumstances (at the margin between taking a measurement and
not taking a measurement) that our theory would yield predictions different
from conventional quantum mechanics postulates.

We now explore this intermediate regime in more details, since it could allow
an experimental test of our theory. The rare hopping between the lobes should
introduce oscillations in the wavefunction with a characteristic spectrum
(which may be detectable through electromagnetic wave emissions if the
particle is charged), which we now derive. Let $b_{t}$ (taking value $0$ or
$1$) denote which of the two identical wavefunction lobes the particle resides
in at time $t$. Let $p_{t}$ denote the total integrated modulus square of the
wavefunction over lobe $1$. Let $r$ denote the particle transition rate from
one lobe to the other, assumed constant, since we consider the regime where
the two lobes persist.

If $b_{t}=0$, then $p_{t}$ decays exponentially with a decay constant
$2\kappa$, while if $b_{t}=1$, then $1-p_{t}$ decays exponentially with the
same decay constant. These two behaviors can be described by the single
equation%
\[
\frac{dp_{t}}{dt}=-2\kappa p_{t}+2\kappa b_{t},
\]
in which $b_{t}$ plays the role of a random driving force. Placing all $p_{t}$
terms on one side and going to frequency ($\omega$) representation (denoted by
tildes), we have%
\begin{equation}
\frac{1}{2\kappa}\left(  i\omega+2\kappa\right)  \tilde{p}_{\omega}=\tilde
{b}_{\omega}\text{ or }\tilde{p}_{\omega}=\frac{2\kappa}{i\omega+2\kappa
}\tilde{b}_{\omega}.\label{eqpb}%
\end{equation}
The real-space autocorrelation function of $b_{t}$ can be determined from $r$
through standard Markov chain techniques on a discretized process and letting
the discretization time step go to zero: $\operatorname*{Covar}\left[
b_{t+\tau},b_{t}\right]  =\frac{1}{4}\exp\left(  -2r\left\vert \tau\right\vert
\right)  $. From this, the corresponding power spectrum is%
\begin{equation}
E\left[  \left\vert \tilde{b}_{\omega}\right\vert ^{2}\right]  =\frac
{r}{\omega^{2}+4r^{2}}.\label{eqbw}%
\end{equation}
It follows from Equations (\ref{eqpb}) and (\ref{eqbw}) that the power
spectrum $E\left[  \left\vert \tilde{p}_{\omega}\right\vert ^{2}\right]  $ is
given by:%
\begin{align*}
E\left[  \left\vert \tilde{p}_{\omega}\right\vert ^{2}\right]   &
=\frac{4\kappa^{2}}{\omega^{2}+4\kappa^{2}}E\left[  \left\vert \tilde
{b}_{\omega}\right\vert ^{2}\right]  =\frac{4\kappa^{2}}{\omega^{2}%
+4\kappa^{2}}\frac{r}{\omega^{2}+4r^{2}}\\
&  \propto\frac{1}{\left(  \frac{\omega}{2\kappa}\right)  ^{2}+1}\cdot\frac
{1}{r}\frac{1}{\left(  \frac{\omega}{2r}\right)  ^{2}+1}.
\end{align*}
This takes the form of a product of two Lorentzian terms whose widths depend,
respectively, on our theory's decay parameter $\kappa$ and on the lobe hopping
rate $r$. By considering different lobe separations (leading to different $r$
but identical $\kappa$), it may be possible to determine $\kappa$. It is
important to note that this electromagnetic signature is very different from
that of typical objective collapse theories. In the latter, any sufficiently
large system should experience wavefunction collapse that generates emissions
(that have so far not been detected). In our theory, emissions only occur when
a system is placed in the exceptional intermediate regime between taking a
measurement and not taking one. Experimental setups enabling this type of test
have been proposed \cite{cavendish:arrival} --- an even more ambitious goal
would be to accomplish this with EPR pairs in order to directly test
superluminal features.

\section{Potential Objections}

\label{appobj}Just like both pilot-wave and objective collapse theories, our
proposal appears to clash with conventional quantum mechanics by introducing a
preferred basis: position eigenstates have a special status compared to
eigenstates of any other observables. This does not present a significant
problem, to the extent that a measurement apparatus typically maps different
values of any observables of interest onto macroscopic states with different
spatial distributions. The latter would initiate the selection of one outcome
based on position and the microscopic system of interest would inherit this
choice through its entanglement with the measurement outcome. This argument
holds even if the microscopic observable of interest is not position.
Furthermore,\ it can be argued that there is some experimental support to the
idea that superpositions of states involving different spatial distributions
are more difficult to obtain. Indeed, superconducting qubits can now routinely
exist in macroscopic (micrometer-size) superpositions of different spin or
momentum states \cite{ezratty:qubitrev}. In contrast, placing a macroscopic
mechanical oscillator in a superposition of states that differ by
$\sim10^{-18}\,$m is considered a remarkable accomplishment
\cite{chu:mechsuper}. Nevertheless, it is possible to modify the theory to
allow for collapse onto a preferred basis other than position --- see Discrete
States section below.

One potential concern with Equation (\ref{eqd2r}) is the fact that the
fictitious particle faces an infinite potential barrier as $\psi\left(
r,t\right)  \longrightarrow0$. This poses no problem for inevitable nodes
introduced by the antisymmetry of the fermionic wavefunctions. Since the
particle has, by construction, an image (under every allowed permutation of
coordinates $\Pi$) on either side of such nodes, there is no need for the
particle to be able to cross the barrier to achieve equilibration. For nodes
that are not forced by antisymmetry, we note that the wavefunction is, in
general, complex-valued and thus a path crossing the origin in the complex
plane is an event of probability zero. In other words, even though nodes are
common in idealized systems, exact zero crossings are rare in realistic
systems where the wavefunction evolves dynamically in low-symmetry environments.

An unusual feature of our approach is the lack of time-reversal symmetry of
our wave equation (\ref{eqmodse}), due to the presence of terms without an
imaginary prefactor in the first derivative expression. A side-effect of this
would be to cause the quantum entropy of a subsystem to increase more slowly
than under the standard Schr\"{o}dinger equation because the wavefunction has
a tendency to localize. This does not necessarily entail a violation of the
second Law of thermodynamics because it merely converts quantum entropy (due
to the subsystem not being in a pure state) into a classical entropy
(reflecting the observer's lack of information regarding the actual state of
the subsystem). From a macroscopic thermodynamic point of view, either
interpretation equally contributes to the macroscopic entropy.

\section{Density Matrix Formulation}

\label{appdens}Our equations of motion can readily be expressed in terms of
the density matrix $\rho$:%
\begin{align}
\frac{d\rho}{dt}  &  =\frac{1}{\mathbf{i}\hbar}\left[  H,\rho\right]
-2\kappa\rho+\kappa\frac{\left[  \bar{\phi}_{r},\rho\right]  _{+}%
}{\left\langle r\right\vert \rho\left\vert r\right\rangle }\label{eqdensmat}\\
\frac{d^{2}r}{dt^{2}}  &  =\frac{\tau}{\mu}\nabla_{r}\ln\left\langle
r\right\vert \rho_{-}\left\vert r\right\rangle ,
\end{align}
where $\bar{\phi}_{r}$ is the operator multiplying by the function $\bar{\phi
}_{r}\left(  x\right)  $ while $\left[  \bar{\phi}_{r},\rho\right]  _{+}%
\equiv\bar{\phi}_{r}\rho+\rho\bar{\phi}_{r}$ and $\rho_{-}$ denotes $\rho$ at
time $t-\Delta t$. Equation (\ref{eqdensmat}) can be derived from the time
differential $d\left(  \left\vert \psi\right\rangle \left\langle
\psi\right\vert \right)  =\left(  d\left\vert \psi\right\rangle \right)
\left\langle \psi\right\vert +\left\vert \psi\right\rangle \left(
d\left\langle \psi\right\vert \right)  $, after observing that $\left\vert
\psi\right\rangle $ is differentiable in our model, so that the quadratic term
$\left(  d\left\vert \psi\right\rangle \right)  \left(  d\left\langle
\psi\right\vert \right)  $\ is negligible. (This behavior is distinct from
that of collapse models \cite{bassi:revcollapse,bassi:objcoluni}, in which
$\left\vert \psi\right\rangle $ is non-differentiable, since it contains a
Wiener process component, and the quadratic term remains, by It\^{o}'s Lemma.)
Next, using this differential to compute $d\left(  \left\vert \psi
\right\rangle \left\langle \psi\right\vert \right)  /dt$ and substituting our
Equation (\ref{eqmodse}) for $d\left\vert \psi\right\rangle /dt$, we have:%
\begin{align*}
\frac{d}{dt}\left\vert \psi\right\rangle \left\langle \psi\right\vert  &
=\left(  \frac{1}{\mathbf{i}\hbar}H-\kappa+\kappa\frac{\bar{\phi}_{r}%
}{\left\vert \psi\left(  r,t\right)  \right\vert ^{2}}\right)  \left\vert
\psi\right\rangle \left\langle \psi\right\vert +\left\vert \psi\right\rangle
\left\langle \psi\right\vert \left(  -\frac{1}{\mathbf{i}\hbar}H-\kappa
+\kappa\frac{\bar{\phi}_{r}}{\left\vert \psi\left(  r,t\right)  \right\vert
^{2}}\right) \\
&  =\frac{1}{\mathbf{i}\hbar}\left[  H,\left\vert \psi\right\rangle
\left\langle \psi\right\vert \right]  -2\kappa\left\vert \psi\right\rangle
\left\langle \psi\right\vert +\kappa\frac{1}{\left\vert \psi\left(
r,t\right)  \right\vert ^{2}}\left(  \bar{\phi}_{r}\left\vert \psi
\right\rangle \left\langle \psi\right\vert +\left\vert \psi\right\rangle
\left\langle \psi\right\vert \bar{\phi}_{r}\right)  ,
\end{align*}
yielding Equation (\ref{eqdensmat}) when one sets $\rho=\left\vert
\psi\right\rangle \left\langle \psi\right\vert $ and notes that $\left\vert
\psi\left(  r,t\right)  \right\vert ^{2}=\left\langle r\right\vert
\rho\left\vert r\right\rangle $.

This density matrix formulation is very helpful to trace out non-positional
degrees of freedom (such as spins) and cleanly obtain an equation of motion
for the positional degrees of freedom only. In conventional quantum mechanics,
one can freely \textquotedblleft trace out\textquotedblright\ the density
matrix to obtain a separate equation of motion for any subsystem, provided
there is no coupling between the subsystem and the environment in the
Hamiltonian $H$. This works even if there is entanglement between the
subsystem and the environment. In our approach, there is also no problem
associated with tracing out the density under these conditions, but the
particle position $r$ cannot always be separated out so easily, a problem
which does not occur when tracing out over non-positional degrees of freedom.
If there is entanglement between the subsystem and the environment, the
trajectory of the particle within the subsystem is still affected by the
environment (through the $\left\langle r\right\vert \rho\left\vert
r\right\rangle $ term), even if there is no coupling through the Hamiltonian.
An independent equation of motion for a subsystem can only be obtained if the
overall density matrix factors as a product of the density matrix of the
system and that of the environment.

\section{Precluding faster-than-light communication}

\label{appftl}We verify the two conditions of Theorem 1 in
\cite{bassi:objcoluni}. We first need to establish that the density matrix
$\rho$ evolves according to a Markov process. In conventional collapse
theories \cite{bassi:revcollapse},\ the values of the random collapse field at
any two different times are completely statistically independent and a
Markovian $\rho$ follows automatically. In our formalism, the role of the
effective random collapse field is played by a smoothed delta function
centered at the Bohmian particle's location $r\left(  t\right)  $. Over a
short time scale, such field unfortunately does not exhibit statistical
independence. Yet, over a time scale larger than the particle's relaxation
time, the current particle's position becomes statistically independent from
the particle's position at a given time in the past. It is known that
measure-preserving dynamical systems (such as our classical particle at
$r\left(  t\right)  $ in a potential $-\tau\ln\left\vert \psi\left(
r,t\right)  \right\vert ^{2}$), typically lead to mixing processes
\cite{halmos:mix} and thus exhibit a characteristic time scale $t_{r}$ for
relaxation. Therefore, if the time evolution of the density matrix cannot be
resolved on a time scale less than some $t_{e}$ (e.g., due to experimental
limitations) and if $t_{e}\gg t_{r}$, then the time evolution of $\rho$ is
indistinguishable from that of a Markov process. In other words, for
sufficiently large sampling time steps $t_{e}\gg t_{r}$, our effective random
field behaves analogously to that of objective collapse theories and so does
$\rho$.

We now turn to the second condition: that the evolution of the statistical
operator $\mathbb{E}\left[  \rho_{t}\right]  \equiv\mathbb{E}\left[
\left\vert \psi_{t}\right\rangle \left\langle \psi_{t}\right\vert \right]  $
is closed and linear, where we added an explicit time-dependence as subscripts
for clarity and $\mathbb{E}$ denotes statistical (not quantum) expectation
values with respect to the random position $r\left(  t\right)  $ of the
particle and the resulting randomness of the wavefunction and density matrix.
Here, closed and linear signifies that, for any two times $s<t$ and whenever
$\sum_{i}\lambda_{i}\left\vert \chi_{i}\right\rangle \left\langle \chi
_{i}\right\vert =\sum_{i}\mu_{i}\left\vert \phi_{i}\right\rangle \left\langle
\phi_{i}\right\vert \equiv\mathbb{E}\left[  \rho_{s}\right]  $, we have
$\sum_{i}\lambda_{i}\mathbb{E}\left[  \left\vert \psi_{t}\right\rangle
\left\langle \psi_{t}\right\vert \mid\psi_{s}=\chi_{i}\right]  =\sum_{i}%
\mu_{i}\mathbb{E}\left[  \left\vert \psi_{t}\right\rangle \left\langle
\psi_{t}\right\vert \mid\psi_{s}=\phi_{i}\right]  $, i.e. expressing the
density matrix as different statistical mixtures of pure states should have no
effect on the dynamics of any observable. To establish\ this, we need to
assume that the typical time scale $t_{c}$ of changes in $\rho$ is also such
that $t_{c}\gg t_{r}$. Then, we also have that the distribution of $r$ for a
given density matrix $\rho$ satisfies $p\left(  r|\rho\right)  \approx
\left\vert \psi\left(  r,t\right)  \right\vert ^{2}$. We can then derive the
time evolution of the expected density matrix using our density matrix
formulation (Equation (\ref{eqdensmat}) in Appendix \ref{appdens}):%
\begin{equation}
\frac{d\mathbb{E}\left[  \rho\right]  }{dt}=\frac{1}{\mathbf{i}\hbar}\left[
H,\mathbb{E}\left[  \rho\right]  \right]  -2\kappa\mathbb{E}\left[
\rho\right]  +\mathbb{E}\left[  \kappa\frac{\left[  \bar{\phi}_{r}%
,\rho\right]  _{+}}{\left\langle r\right\vert \rho\left\vert r\right\rangle
}\right]  \label{eqexprho}%
\end{equation}
where the expectations of the first three terms go through the operators by
linearity (under standard regularity conditions on $\rho$). The last term
requires more care:%
\begin{align*}
\mathbb{E}\left[  \kappa\frac{\left[  \bar{\phi}_{r},\rho\right]  _{+}%
}{\left\langle r\right\vert \rho\left\vert r\right\rangle }\right]   &
=\mathbb{E}\left[  \mathbb{E}\left[  \kappa\frac{\left[  \bar{\phi}_{r}%
,\rho\right]  _{+}}{\left\langle r\right\vert \rho\left\vert r\right\rangle
}\mid\rho\right]  \right]  =\mathbb{E}\left[  \int\kappa\frac{\left[
\bar{\phi}_{r},\rho\right]  _{+}}{\left\langle r\right\vert \rho\left\vert
r\right\rangle }p\left(  r|\rho\right)  dr\right]  \\
&  =\mathbb{E}\left[  \int\kappa\frac{\left[  \bar{\phi}_{r},\rho\right]
_{+}}{\left\langle r\right\vert \rho_{-}\left\vert r\right\rangle
}\left\langle r\right\vert \rho_{-}\left\vert r\right\rangle dr\right]
=\mathbb{E}\left[  \int\kappa\left[  \bar{\phi}_{r},\rho\right]
_{+}dr\right]  \\
&  =\mathbb{E}\left[  \kappa\left[  \int\bar{\phi}_{r}dr,\rho\right]
_{+}\right]  =\mathbb{E}\left[  \kappa\left[  1,\rho\right]  _{+}\right]
=2\kappa\mathbb{E}\left[  \rho\right]  ,
\end{align*}
where we have used, in turn, (i) iterated expectations, (ii) the definition of
conditional expectation in terms of a conditional density $p\left(
r|\rho\right)  $ of $r$ given $\rho$, (iii) the assumption that the
equilibration rate $t_{r}^{-1}$ of the distribution of $r$ is fast relative to
the rate $t_{c}^{-1}$ of change in $\rho$, so that $p\left(  r|\rho\right)
=\left\vert \psi\left(  r,t\right)  \right\vert ^{2}=\left\langle r\right\vert
\rho\left\vert r\right\rangle =\left\langle r\right\vert \rho_{-}\left\vert
r\right\rangle $, (iv) the observation that the only remaining dependence on
$r$ lies in $\bar{\phi}_{r}$, (v) the facts that $\int\bar{\phi}_{r}dr=1$ and
(vi) that $1\rho=\rho1=\rho$. It follows that Equation (\ref{eqexprho})
reduces to
\begin{equation}
\frac{d\mathbb{E}\left[  \rho\right]  }{dt}=\frac{1}{\mathbf{i}\hbar}\left[
H,\mathbb{E}\left[  \rho\right]  \right]  ,\label{eqexprhosimp}%
\end{equation}
so that the expected density matrix $\mathbb{E}\left[  \rho\right]  $ follows
a standard Liouville-von Neumann equation, just like the density matrix in
conventional quantum mechanics. This implies that the time evolution of the
expected density matrix is indeed independent of the choice of pure-state
mixture used to represent it.

Having verified the assumptions of Theorem 1 in \cite{bassi:objcoluni}, we
have thus established that if the relaxation time $t_{r}$ for the particle
position is significantly less than both the time scale $t_{c}$ of changes in
the density matrix and the experimentally accessible time resolution $t_{e}$,
then no faster-than-light communication is possible.

Let us now consider the opposite limit. Consider a system initially such that
the distribution of $r$ had reached equilibrium and that then undergoes a
measurement operation in some basis $\left\vert \phi_{i}\right\rangle $. Then,
the wavefunction will exhibit well-separated lobes that can trap the particle,
and this momentary loss of ergodicity indicates that the relaxation time
$t_{r}$ becomes long in that case, until the particle equilibrates once again
within the randomly selected lobe. We have argued that, in this case, Born's
rule is satisfied in the sense that an initial state $\left\vert
\psi\right\rangle $ evolves to the state $\left\vert \phi_{i}\right\rangle $
with probability $\left\vert \left\langle \phi_{i}|\psi\right\rangle
\right\vert ^{2}$, or, in density matrix formulation, the pure nonrandom state
$\rho=\mathbb{E}\left[  \rho\right]  =\left\vert \psi\right\rangle
\left\langle \psi\right\vert $ evolves to $\mathbb{E}\left[  \rho\right]
=\sum_{i}\left\vert \phi_{i}\right\rangle \left\vert \left\langle \phi
_{i}|\psi\right\rangle \right\vert ^{2}\left\langle \phi_{i}\right\vert
=\sum_{i}\left\vert \phi_{i}\right\rangle \left\langle \phi_{i}|\rho|\phi
_{i}\right\rangle \left\langle \phi_{i}\right\vert $, an expression which is
linear in the initial state $\rho$ and that does not depend on earlier states.
The assumptions of Theorem 1 in \cite{bassi:objcoluni} are once again
satisfied and faster-than-light communication is prevented.

In short, the expected time evolution of the density matrix under our dynamics
either reduces to the standard Liouville-von Neumann equation (when ergodicity
holds), or reduces to standard wavefunction collapse (when ergodicity fails).
Either way, the density matrix evolution remains Markovian and the time
evolution of its expectation remains linear, thus precluding faster-than-light
communication in those regimes. The only way faster-than-light communication
could possibly manifest itself would be in an intermediate regime where the
relaxation time of the distribution of $r$ is neither short nor long. That is,
one would need to create a wavefunction with lobes that are somewhat separated
but not completely, so that $r$ never has the time to fully equilibrate before
it jumps to another lobe. In other words, one would need a way to somehow
interrupt the measurement process midway to make another measurement or set up
conditions at the margin between taking a measurement or not (as explored in
Appendix \ref{apptrans}). Hence, the conditions under which faster-than-light
communication might be detected are arguably difficult to create.\ The
possibility that non-equilibrium Bohmian particles could lead to a breakdown
of locality has been suggested before \cite{valentini:noneq}.

\section{Discrete States}

\label{appdisc}A potentially useful extension of our theory would be to
include systems with discrete states where there is no natural definition of
position and momentum. This would allow for wavefunction collapse in spin
systems or fermionic quantum fields, for instance, without resorting to the
argument that the collapse only occurs when such systems are coupled with a
macroscopic position-based pointer.

In some sense, this task is far easier to do with our approach than with
standard pilot-wave theory, because our equation of motion for the particle
position $r$ does not rely on the definition of a momentum operator. A natural
discrete formulation of our modified Schr\"{o}dinger equation would be:%
\[
\frac{d\psi\left(  t\right)  }{dt}=\frac{1}{\mathbf{i}\hbar}H\psi\left(
t\right)  -\kappa\psi\left(  t\right)  +\kappa\frac{\delta_{r\left(  t\right)
}}{\left\vert \psi_{r\left(  t\right)  }\left(  t\right)  \right\vert ^{2}%
}\psi\left(  t\right)
\]
where the wavefunction $\psi\left(  t\right)  $ is represented by a vector
with elements $\psi_{r}\left(  t\right)  $, while $r\left(  t\right)  $
represents the current discrete state of the Bohmian particle at time $t$. The
localization function $\bar{\phi}_{r}$ just becomes a Kronecker delta
$\delta_{r}$. It is not clear how to obtain a deterministic equation of motion
for $r\left(  t\right)  $, but one could postulate a stochastic evolution
equation (paralleling \cite{vink:bohmdisc}) that specifies the probability of
transition $P_{r^{\prime}r}$ from a state $r$ to another state $r^{\prime}$
per unit time. The transition probability matrix $P_{r^{\prime}r}$ must
satisfy the condition that it admits an attractive equilibrium distribution of
$r$ that matches $\left\vert \psi_{r}\right\vert ^{2}$. The theory of Monte
Carlo sampling provides many convenient valid choices, for instance:%
\[
P_{r^{\prime}r}\propto\left(  1+\frac{\left\vert \psi_{r}\right\vert ^{2}%
}{\left\vert \psi_{r^{\prime}}\right\vert ^{2}}\right)  ^{-1}K_{r^{\prime}r}%
\]
for some given function $K_{r^{\prime}r}$ satisfying $K_{r^{\prime}%
r}=K_{rr^{\prime}}$. The function $K_{r^{\prime}r}$ essentially specifies
which pairs of states $r$, $r^{\prime}$ are considered in proximity.
Presumably, $K_{r^{\prime}r}$ should be larger for transitions in which
$r^{\prime}$ differs from $r$ in the state of a few particles (say, one spin
value $\left\vert \uparrow\right\rangle $ or $\left\vert \downarrow
\right\rangle $). For such choice, joint states of the form $\left(
\left\vert \uparrow\right\rangle +\left\vert \downarrow\right\rangle \right)
\otimes\cdots\otimes\left(  \left\vert \uparrow\right\rangle +\left\vert
\downarrow\right\rangle \right)  $ experience very little trapping while
entangled states of the form $\left\vert \uparrow\right\rangle \otimes
\cdots\otimes\left\vert \uparrow\right\rangle +\left\vert \downarrow
\right\rangle \otimes\cdots\otimes\left\vert \downarrow\right\rangle $ would
experience the most trapping, because it would take many single-spin flips to
go from one of the pure states to the other. Absent this preference for small
changes, there is little possibility of trapping and we would essentially
recover Schr\"{o}dinger's equation. We leave a more precise specification of
$K_{r^{\prime}r}$ for future work.

\bibliographystyle{elsarticle-num} 
\bibliography{meas.bib}

\begin{thebibliography}{10}
\expandafter\ifx\csname url\endcsname\relax
  \def\url#1{\texttt{#1}}\fi
\expandafter\ifx\csname urlprefix\endcsname\relax\def\urlprefix{URL }\fi
\expandafter\ifx\csname href\endcsname\relax
  \def\href#1#2{#2} \def\path#1{#1}\fi

\bibitem{cavalcanti:persp}
E.~G. Cavalcanti, R.~Chaves, F.~Giacomini, Y.-C. Liang, Fresh perspectives on
  the foundations of quantum physics, Nat. Rev. Phys. 5 (2023) 323.
\newblock \href {https://doi.org/10.1038/s42254-023-00586-z}
  {\path{doi:10.1038/s42254-023-00586-z}}.

\bibitem{wiseman:wigfri}
K.-W. Bong, A.~Utreras-Alarc{\'o}n, F.~Ghafari, Y.-C. Liang, N.~Tischler, E.~G.
  Cavalcanti, G.~J. Pryde, H.~M. Wiseman, A strong no-go theorem on the
  wigner's friend paradox, Nat. Phys. 16 (2020) 1199.
\newblock \href {https://doi.org/10.1038/s41567-020-0990-x}
  {\path{doi:10.1038/s41567-020-0990-x}}.

\bibitem{bassi:exptest}
S.~Donadi, K.~Piscicchia, C.~Curceanu, L.~Di{\'o}si, M.~Laubenstein, A.~Bassi,
  Underground test of gravity-related wave function collapse, Nat. Phys. 17
  (2021) 74.
\newblock \href {https://doi.org/10.1038/s41567-020-1008-4}
  {\path{doi:10.1038/s41567-020-1008-4}}.

\bibitem{spekkens:contex}
J.~H. Selby, D.~Schmid, E.~Wolfe, A.~B. Sainz, R.~Kunjwal, R.~W. Spekkens,
  Contextuality without incompatibility, Phys. Rev. Lett. 130 (2023) 230201.
\newblock \href {https://doi.org/10.1103/PhysRevLett.130.230201}
  {\path{doi:10.1103/PhysRevLett.130.230201}}.

\bibitem{zurek:deco}
W.~H. Zurek, Environment-induced superselection rules, Phys. Rev. D 26 (1982)
  1862.
\newblock \href {https://doi.org/10.1103/PhysRevD.26.1862}
  {\path{doi:10.1103/PhysRevD.26.1862}}.

\bibitem{schloss:deco}
M.~Schlosshauer, Decoherence, the measurement problem, and interpretations of
  quantum mechanics, Rev. Mod. Phys. 76 (2005) 1267.
\newblock \href {https://doi.org/10.1103/RevModPhys.76.1267}
  {\path{doi:10.1103/RevModPhys.76.1267}}.

\bibitem{valentini:bohmrev}
A.~Valentini, De {Broglie-Bohm} quantum mechanics, in: R.~Szabo, M.~Bojowald
  (Eds.), Encyclopedia of Mathematical Physics, 2nd Edition, Academic Press,
  Oxford, 2025, pp. 24--41.
\newblock \href {https://doi.org/10.1016/B978-0-323-95703-8.00264-0}
  {\path{doi:10.1016/B978-0-323-95703-8.00264-0}}.

\bibitem{bohm:hidvar12}
D.~Bohm, A suggested interpretation of the quantum theory in terms of hidden
  variables, {I} and {II}, Phys. Rev. 85 (1952) 166.
\newblock \href {https://doi.org/10.1103/PhysRev.85.166}
  {\path{doi:10.1103/PhysRev.85.166}}.

\bibitem{broglie:pilot}
L.~de~Broglie, La m{\'{e}}canique ondulatoire et la structure atomique de la
  mati{\`{e}}re et du rayonnement, J. Phys. Radium 8 (1927) 225.
\newblock \href {https://doi.org/10.1051/jphysrad:0192700805022500}
  {\path{doi:10.1051/jphysrad:0192700805022500}}.

\bibitem{durr:bohm}
D.~D{\"{u}}rr, S.~Teufel, Bohmian mechanics, Springer-Verlag, Heidelberg, 2009.

\bibitem{carroll:many}
S.~M. Carroll, C.~T. Sebens, Many worlds, the {b}orn rule, and self-locating
  uncertainty, in: D.~Struppa, J.~Tollaksen (Eds.), Quantum theory: A two-time
  success story: Yakir Aharonov Festschrift, Springer, Milan, 2014, pp.
  157--169.
\newblock \href {https://doi.org/10.1007/978-88-470-5217-8\_10}
  {\path{doi:10.1007/978-88-470-5217-8\_10}}.

\bibitem{everett:relative}
H.~Everett, ``{R}elative state'' formulation of quantum mechanics, Rev. Mod.
  Phys. 29 (1957) 454.
\newblock \href {https://doi.org/10.1103/RevModPhys.29.454}
  {\path{doi:10.1103/RevModPhys.29.454}}.

\bibitem{fuchs:qbism}
C.~A. Fuchs, R.~Schack, Quantum-{B}ayesian coherence, Rev. Mod. Phys. 85 (2013)
  1693.
\newblock \href {https://doi.org/10.1103/RevModPhys.85.1693}
  {\path{doi:10.1103/RevModPhys.85.1693}}.

\bibitem{bassi:revcollapse}
A.~Bassi, K.~Lochan, S.~Satin, T.~P. Singh, H.~Ulbricht, Models of
  wave-function collapse, underlying theories, and experimental tests, Rev.
  Mod. Phys. 85 (2013) 471.
\newblock \href {https://doi.org/10.1103/RevModPhys.85.471}
  {\path{doi:10.1103/RevModPhys.85.471}}.

\bibitem{penrose:gravqm}
R.~Penrose, On the gravitization of quantum mechanics 1: Quantum state
  reduction, Found. Phys. 44 (2014) 557.
\newblock \href {https://doi.org/10.1007/s10701-013-9770-0}
  {\path{doi:10.1007/s10701-013-9770-0}}.

\bibitem{diodi:grav}
L.~Diosi, Models for universal reduction of macroscopic quantum fluctuations,
  Phys. Rev. A 40 (1989) 1165.
\newblock \href {https://doi.org/10.1103/PhysRevA.40.1165}
  {\path{doi:10.1103/PhysRevA.40.1165}}.

\bibitem{schonfeld:cloud}
J.~F. Schonfeld, Physical origins and limitations of canonical quantum
  measurement behavior, Tech. Rep. 2505.00716, {ArXiv} (2025).
\newblock \href {https://doi.org/10.48550/arXiv.2505.00716}
  {\path{doi:10.48550/arXiv.2505.00716}}.

\bibitem{balian:dynmeas}
A.~E. Allahverdyan, R.~Balian, T.~M. Nieuwenhuizen, Understanding quantum
  measurement from the solution of dynamical models, Physics Reports 525~(1)
  (2013) 1.
\newblock \href {https://doi.org/10.1016/j.physrep.2012.11.001}
  {\path{doi:10.1016/j.physrep.2012.11.001}}.

\bibitem{melkikh:nlqmeas}
A.~V. Melkikh, Nonlinearity of quantum mechanics and solution of the problem of
  wave function collapse, Commun. Theor. Phys. 64 (2015) 47.
\newblock \href {https://doi.org/10.1088/0253-6102/64/1/47}
  {\path{doi:10.1088/0253-6102/64/1/47}}.

\bibitem{laloe:attract}
F.~Lalo{\"e}, Quantum collapse dynamics with attractive densities, Phys. Rev. A
  99 (2019) 052111.
\newblock \href {https://doi.org/10.1103/PhysRevA.99.052111}
  {\path{doi:10.1103/PhysRevA.99.052111}}.

\bibitem{barandes:sqcorr}
J.~Barandes, The stochastic-quantum correspondence, Philosophy of Physics 3
  (2025) 8.
\newblock \href {https://doi.org/10.31389/pop.186}
  {\path{doi:10.31389/pop.186}}.

\bibitem{wiseman:disguise}
A.~Tilloy, H.~M. Wiseman, Non-markovian wave-function collapse models are
  {B}ohmian-like theories in disguise, Quantum 5 (2021) 594.
\newblock \href {https://doi.org/10.22331/q-2021-11-29-594}
  {\path{doi:10.22331/q-2021-11-29-594}}.

\bibitem{schlosshauer:snapshot}
M.~Schlosshauer, J.~Kofler, A.~Zeilinger, A snapshot of foundational attitudes
  toward quantum mechanics, Stud. Hist. Phil. Mod. Phys. 44 (2013) 222.
\newblock \href {https://doi.org/10.1016/j.shpsb.2013.04.004}
  {\path{doi:10.1016/j.shpsb.2013.04.004}}.

\bibitem{struyve:bohmqft}
W.~Struyve, Pilot-wave theory and quantum fields, Rep. Prog. Phys. 73 (2010)
  106001.
\newblock \href {https://doi.org/10.1088/0034-4885/73/10/106001}
  {\path{doi:10.1088/0034-4885/73/10/106001}}.

\bibitem{mcquarrie:statmech}
D.~A. McQuarrie, Statistical Mechanics, University Science Books, New York,
  2000.

\bibitem{villani:hthm}
L.~Desvillettes, C.~Villani, On the trend to global equilibrium for spatially
  inhomogeneous kinetic systems: The boltzmann equation, Invent. math. 159
  (2005) 245.
\newblock \href {https://doi.org/10.1007/s00222-004-0389-9}
  {\path{doi:10.1007/s00222-004-0389-9}}.

\bibitem{valentini:noneq}
A.~Valentini, Signal-locality, uncertainty, and the sub-quantum {H}-theorem,
  {II}, Physics Letters A 158 (1991) 1.
\newblock \href {https://doi.org/10.1016/0375-9601(91)90330-B}
  {\path{doi:10.1016/0375-9601(91)90330-B}}.

\bibitem{vineyard:tst}
G.~H. Vineyard, Frequency factors and isotope effects in solid state rate
  processes, J. Phys. Chem. Solids 3 (1957) 121.
\newblock \href {https://doi.org/10.1016/0022-3697(57)90059-8}
  {\path{doi:10.1016/0022-3697(57)90059-8}}.

\bibitem{EPR:org}
A.~Einstein, B.~Podolsky, N.~Rosen, Can quantum-mechanical description of
  physical reality be considered complete?, Phys. Rev. 47 (1935) 777.
\newblock \href {https://doi.org/10.1103/PhysRev.47.777}
  {\path{doi:10.1103/PhysRev.47.777}}.

\bibitem{zanghi:bm_grw}
V.~Allori, S.~Goldstein, R.~Tumulka, N.~Zangh{\`\i}, On the common structure of
  {B}ohmian mechanics and the {G}hirardi-{R}imini-{W}eber theory, The British
  Journal for the Philosophy of Science 59 (2008) 353.
\newblock \href {https://doi.org/10.1093/bjps/axn012}
  {\path{doi:10.1093/bjps/axn012}}.

\bibitem{darrow:2way}
D.~Darrow, J.~W.~M. Bush, Revisiting de {B}roglie's double-solution pilot-wave
  theory with a {L}orentz-covariant lagrangian framework, Symmetry 16 (2024)
  149.
\newblock \href {https://doi.org/10.3390/sym16020149}
  {\path{doi:10.3390/sym16020149}}.

\bibitem{bollinger:linqm}
J.~J. Bollinger, D.~J. Heinzen, W.~M. Itano, S.~L. Gilbert, D.~J. Wineland,
  Test of the linearity of quantum mechanics by rf spectroscopy of the
  {$^9$Be$^+$} ground state, Phys. Rev. Lett. 63 (1989) 1031.
\newblock \href {https://doi.org/10.1103/PhysRevLett.63.1031}
  {\path{doi:10.1103/PhysRevLett.63.1031}}.

\bibitem{polchinski:nlqmbug}
J.~Polchinski, Weinberg's nonlinear quantum mechanics and the
  {E}instein-{P}odolsky-{R}osen paradox, Phys. Rev. Lett. 66 (1991) 397.
\newblock \href {https://doi.org/10.1103/PhysRevLett.66.397}
  {\path{doi:10.1103/PhysRevLett.66.397}}.

\bibitem{bassi:objcoluni}
A.~Bassi, D.~D{\"u}rr, G.~Hinrichs, Uniqueness of the equation for quantum
  state vector collapse, Phys. Rev. Lett. 111 (2013) 210401.
\newblock \href {https://doi.org/10.1103/PhysRevLett.111.210401}
  {\path{doi:10.1103/PhysRevLett.111.210401}}.

\bibitem{zanghi:bohmrel}
D.~D{\"{u}}rr, S.~Goldstein, T.~Norsen, W.~Struyve, N.~Zangh{\`{i}}, Can
  {B}ohmian mechanics be made relativistic?, Proc. R. Soc. A 470 (2014)
  20130699.
\newblock \href {https://doi.org/10.1098/rspa.2013.0699}
  {\path{doi:10.1098/rspa.2013.0699}}.

\bibitem{tumulka:grwrel}
R.~Tumulka, A relativistic version of the {G}hirardi-{R}imini-{W}eber model, J.
  of Stat. Phys. 125 (2006) 821.
\newblock \href {https://doi.org/10.1007/s10955-006-9227-3}
  {\path{doi:10.1007/s10955-006-9227-3}}.

\bibitem{kovachy:qmetrescale}
T.~Kovachy, P.~Asenbaum, C.~Overstreet, C.~A. Donnelly, S.~M. Dickerson,
  A.~Sugarbaker, J.~M. Hogan, M.~A. Kasevich, Quantum superposition at the
  half-metre scale, Nature 528 (2015) 530.
\newblock \href {https://doi.org/10.1038/nature16155}
  {\path{doi:10.1038/nature16155}}.

\bibitem{cavendish:arrival}
A.~A. Rafsanjani, W.~Cavendish, The arrival position problem in quantum
  mechanics, Tech. Rep. 2607.03538, {ArXiv} (2026).
\newblock \href {https://doi.org/10.48550/arXiv.2607.03538}
  {\path{doi:10.48550/arXiv.2607.03538}}.

\bibitem{ezratty:qubitrev}
O.~Ezratty, Perspective on superconducting qubit quantum computing, Eur. Phys.
  J. A 59 (2023) 94.
\newblock \href {https://doi.org/10.1140/epja/s10050-023-01006-7}
  {\path{doi:10.1140/epja/s10050-023-01006-7}}.

\bibitem{chu:mechsuper}
M.~Bild, M.~Fadel, Y.~Yang, U.~von L{\"u}pke, P.~Martin, A.~Bruno, Y.~Chu,
  Schr{\"{o}}dinger cat states of a 16-microgram mechanical oscillator, Science
  380 (2023) 274.
\newblock \href {https://doi.org/10.1126/science.adf7553}
  {\path{doi:10.1126/science.adf7553}}.

\bibitem{halmos:mix}
P.~R. Halmos, In general a measure preserving transformation is mixing, Annals
  of Mathematics 45 (1944) 786.
\newblock \href {https://doi.org/10.2307/1969304} {\path{doi:10.2307/1969304}}.

\bibitem{vink:bohmdisc}
J.~C. Vink, Quantum mechanics in terms of discrete beables, Phys. Rev. A 48
  (1993) 1808.
\newblock \href {https://doi.org/10.1103/PhysRevA.48.1808}
  {\path{doi:10.1103/PhysRevA.48.1808}}.

\end{thebibliography}

\end{document}